\documentclass[twoside,twocolumn,english,superscriptaddress,nofootinbib,preprintnumbers, aps]{revtex4}
\usepackage{color}
\usepackage{amsmath}
\usepackage{graphicx}
\usepackage{amssymb}
\usepackage{esint}
\usepackage[hyperfootnotes=true,unicode=true, 
 bookmarks=true,bookmarksnumbered=false,bookmarksopen=false,
 breaklinks=false,pdfborder={0 0 1},backref=false,colorlinks=true]
 {hyperref}

\makeatletter

\@ifundefined{textcolor}{}
{%
 \definecolor{BLACK}{gray}{0}
 \definecolor{WHITE}{gray}{1}
 \definecolor{RED}{rgb}{1,0,0}
 \definecolor{GREEN}{rgb}{0,1,0}
 \definecolor{BLUE}{rgb}{0,0,1}
 \definecolor{CYAN}{cmyk}{1,0,0,0}
 \definecolor{MAGENTA}{cmyk}{0,1,0,0}
 \definecolor{YELLOW}{cmyk}{0,0,1,0}
 }

\usepackage{hyperref}
\hypersetup{
    colorlinks,%
    citecolor=blue,%
    filecolor=blue,%
    linkcolor=blue,%
    urlcolor=blue
}

\makeatother

\begin{document}

\title{R-Symmetry and Gravitino Abundance}

\author{Ioannis Dalianis}
\email{Ioannis.Dalianis@fuw.edu.pl}
\affiliation{Institute of Theoretical Physics, Faculty of  Physics \\
University of Warsaw, ul. Ho\.za 69, Warsaw, Poland}


\begin{abstract}
Supersymmetry breaking sectors generically have an approximate global $U(1)$ $R$-symmetry. Thermal effects tend to restore the $R$-symmetry in the reheated early universe. We revisit the gravitino generation from the thermal plasma and we argue that an $R$-symmetric phase suppresses the gaugino masses and thereby the production of the helicity $\pm 1/2$ gravitino component. For reasonable values of the hidden sector parameters the gravitino can account for the dark matter of the universe with a relic abundance characterized by a remarkable insensitivity to the reheating temperature.

\end{abstract}


\maketitle

\paragraph*{Introduction.}

A generic feature of the supersymmetry breaking sectors is the presence of an exact or approximate continuous global $U(1)$ symmetry called $R$-symmetry. As pointed out at \cite{Nelson:1993nf} a continuous $R$-symmetry is a necessary condition for supersymmetry breaking and a spontaneously broken $R$-symmetry a sufficient condition if the superpotential is a generic function of the fields. Hence, if there is a supersymmetry breaking vacuum an $R$-symmetry, exact or approximate, must be generally expected \cite{Intriligator:2007py}.
Further support for the  models characterized by an $R$-symmetry comes from the requirement that the cosmological constant vanishes in the vacuum \cite{Dine:2009swa} or the necessity to explain large hierarchies \cite{Kappl:2008ie}. 
However, $R$-symmetry must be broken in order that Majorana gaugino masses are generated.

In the thermalized early universe the $R$-symmetry, as we will discuss, although broken at low energies can be restored at high temperatures. 
This entails that the gaugino masses had been $R$-suppressed, a  fact that leads us to reconsider the relic gravitino abundance. 
Here, we will derive a new expression for the relic density of the thermally produced stable gravitinos which does not have the conventional linear dependence on the reheating temperature \cite{Ellis:1984eq,Moroi:1993mb,Bolz:2000fu, Pradler:2006qh, Rychkov:2007uq}. On the contrary, it depends on a temperature characteristic of the $R$-symmetric (exact or approximate) thermal phase. This result implies a significant relaxation of the reheating temperature bound in order that  the
dark matter constraint $\Omega_{\tilde{G}}h^2 \lesssim 0.11$ \cite{Komatsu:2010fb} is satisfied.
Furthermore,  it commends that the gravitino with mass in the range ${\cal O}$(MeV)-${\cal O}$(GeV) can generically account for the dominant dark matter component of the universe. 

\paragraph*{ $R$-Symmetry and Gravitino Production.}
In gauge mediation \cite{Giudice:1998bp} the gaugino masses $m_\lambda$ arise at one-loop 
from the $R$-violating operator $\int d^2\theta \ln X W^\alpha W_\alpha + \text{h.c.}$ where $X$ the $R$-charged spurion superfield that breaks supersymmetry.
When the $U(1)_R$ symmetry is a symmetry of the vacuum soft masses for Majorana gauginos are not allowed; see \cite{Benakli:2010gi} for a different scenario.

A global $U(1)_R$ symmetry broken at zero temperature can, in principle, be restored at finite temperatures. The phase transition from an $R$-violating phase to an $R$-preserving one has important implications on the thermal production rate for the helicity $\pm 1/2$ gravitinos: gauginos become massless hence, 
the reaction rate that generate gravitinos is not $(m_\lambda/m_{\tilde{G}})^2$ enhanced but is simply given by the  $\gamma_{sc}\sim T^6/M^2_P$ factor. 
Let us call the temperature of the transition to the $R$-symmetric phase $T_0$. Then for temperatures larger than $T_0$ the production of gravitinos is dominated by the helicity $\pm 3/2$ component i.e. for light gravitinos it is substantially eliminated. Only for temperatures $T<T_0$ the helicity $\pm 1/2$ (Goldstino) production rate is recovered and, unless the $\pm 3/2$ component is overpopulated the yield $Y_{\tilde{G}}$ is dominated by the $T_0$ and \itshape not \normalfont from the reheating temperature $T_{rh}$.

The scalar soft masses that arise at two loops from the operator $\int d^4 \theta \ln (X^\dagger X) Q^\dagger e^V Q$ do not violate the $R$-symmetry. However, the contribution of sfermions to the Goldstino production is subleading whence, we can focus only at those $2\rightarrow 2$ reactions that include 
 a member of gauge supermultiplets and ignore those with chiral supermultiplets altogether. 

Gravitinos can be also produced from the thermal plasma due to the top Yukawa coupling, an effect considered in \cite{Rychkov:2007uq}. Like the gaugino masses, the $A$-terms require interactions which violate the  $U(1)_R$ symmetry and therefore, we expect them to be suppressed as well at high temperatures.

\paragraph*{Thermal Restoration of the $U(1)_R$ Symmetry.}

Let us consider the standard paradigm of gauge mediation:  
\begin{equation} \label{min-3}
W=FX+\lambda X\phi \bar{\phi}.
\end{equation}
The $\phi$ and $\bar{\phi}$ are the messenger fields. This superpotential  yields universal soft masses, i.e. sfermions and gauginos have masses of the same order. 
The (\ref{min-3}) is $R$-symmetric and a requirement for universal masses is the breaking of the $R$-symmetry at the vacuum. 
At tree level the $X$-direction is a pseudoflat direction with non-zero $F$-term. 
Everywhere along the pseudoflat direction the $R$-symmetry is spontaneously broken apart from the origin where the $R$-symmetry is conserved.  Perturbative corrections lift the flatness. 
The integrated out degrees of freedom also induce a correction $\delta K$ to the canonical  K\"ahler  
\begin{equation}
\delta K=|X|^4/\Lambda^2\, 
\end{equation}
where $\Lambda$ the cut-off of the theory. Thereby, there is the quadratic term $4F^2|X|^2/\Lambda^2$ at the zero temperature potential. At finite temperature the messenger fields induce a thermal mass on the spurion of the form $T^2\lambda^2|X|^2 N/4$  where $N$ are the complex degrees of freedom for the messenger fields $\phi$ ($\bar{\phi}$) considered to be in a fundamental (anti-fundamental) representation of a GUT group.  Hence, for models that exhibit spontaneous $R$-symmetry breaking the critical temperature for the second order phase transition towards the broken $R$ phase is \cite{Dalianis:2010yk}
\begin{equation} \label{TR}
T_0=\frac{4}{\sqrt{N}} \frac{F}{\lambda \Lambda}\,.
\end{equation}
Models with a K\"ahler potential of the form $ K=|X|^2+|X|^4/\Lambda^2 - \epsilon |X|^6/\Lambda^4$ where $\epsilon$ a positive factor are some examples \cite{Lalak:2008bc, Dalianis:2010yk}. This K\"ahler may originate from a massive integrated out sector that breaks the $R$-symmetry spontaneously. An example is the O'Raifeartaigh-type models proposed in \cite{Shih:2007av} that contain fields with $R$-charge $R \neq 0,\,2$ or the model discussed in \cite{Dimopoulos:1997ww}.

In a thermalized universe it is the messenger fields $\phi$, $\bar{\phi}$ that control the thermal average value of the $R$-charged spurion $X$-field and thus, the thermal restoration of the $R$-symmetry.
When the $R$-symmetry is thermally restored the thermal average value is $X(T)=0$. We can define the $R$-breaking parameter $b_R(T)$
\begin{equation}\label{bR}
b_R(T) \equiv \frac{X(T)}{X_0}\,.
\end{equation}
For a second order phase transitions, like the one discussed above, it takes the approximate values $b_R(T>T_0)=0$ and $b_R(T<T_0)=1$.

Let us consider now theories that break the $R$-symmetry explicitly. For instance the superpotential (\ref{min-3}) may be supplemented by a constant term $\delta W=W_0$ that cancels the cosmological constant or an $R$-violating messenger mass term $\delta W =M \phi\bar{\phi}$. We can assume also here a corrected K\"ahler of the form $K=|X|^2-|X|^4/\Lambda^2$. The $R$-violating terms result in a linear to $X$ term at the zero temperature scalar potential for $X$
that we write in the general form
\begin{equation}
V=V_0 +\alpha_1 (X+X^\dagger) +\alpha_2 |X|^2 + {\cal O}(X^3)\, ,
\end{equation}
where $\alpha_2=4F^2/\Lambda^2$. The $R$-violating vev of the $X$ field is $X_0=-\alpha_1/\alpha_2$. At finite temperature $T<\Lambda$ the thermal average value for the $R$-charged $X$ field over the zero temperature vev $X_0$  reads 
\begin{equation} \label{XT}
b_R(T)\equiv \frac{X(T)}{X_0}\simeq \frac{\alpha_2}{\alpha_2+\frac{N}{4}\lambda^2 T^2}\,.
\end{equation}
This is recast to 
\begin{equation} \label{bRa}
b_R(T) = \frac{1}{1+\left(\frac{T}{T_0} \right)^2}\,
\end{equation}
where the  $T_0$ is given by the eq. (\ref{TR}). Here, at $T_0$ there is no any second order phase transition. There is never a barrier and the symmetry breaking occurs via a graduate increase of the mean value of the $X(T)$ field.  At high temperatures the  $R$-symmetry tends to be restored i.e. $b_R(T)$ tends to zero. Indeed, for high enough temperatures the $R$-violating terms $W_0$ and $M\phi\bar{\phi}$ are negligible and an approximate $R$-symmetry restoration takes place. 
We note that for the case of constant term $W_0$ the zero temperature vev is $X_0 =W_0\Lambda^2/2FM^2_P$ \cite{Kitano:2006wz} and for the messenger mass case, after the translation $X\rightarrow M/\lambda-X$,  is $X_0=M/\lambda$.
Obviously, when $T\rightarrow 0$ the $R$-symmetry breaking scale takes its maximum value, i.e. the zero temperature one, and when $T \rightarrow \infty $ the $R$-symmetry is restored. In other words, the $b_R(T)$ parameterizes the $R$-symmetry breaking scale at finite temperature with respect to the zero temperature breaking scale.

\paragraph*{Gravitino Thermal Production Revisited.}

The thermal tendency to restore the $R$-symmetry has important consequences. When the $R$-symmetry is fully restored the gaugino masses vanish; for the case of approximate $R$-symmetry restoration the gaugino masses are accordingly suppressed. The expected suppression is of the form $m_\lambda(T) \sim b_R(T)m_\lambda$.
Here we consider:
\begin{equation} \label{ratio-9}
\frac{m_\lambda(T)}{m_\lambda} =\frac{X(T)}{X_0}= b_R(T)\, 
\end{equation}
where $m_\lambda$ the zero temperature gaugino mass. Therefore, the gravitino cross section from the scattering processes off thermal radiations, $\left\langle \Sigma_{\tilde{G}} v\right\rangle$, becomes thermally dependent. 
Considering the dominant QCD $2\rightarrow 2$ processes \cite{Bolz:2000fu} the cross section is modulated by the $b_R(T)$ factor
\begin{equation} \label{crossT-9}
\Sigma_{\tilde{G}}(p, T) \propto \frac{g^2}{M^2_P} \left(1+b^2_R(T)\frac{m^2_{\tilde{g}}}{3m^2_{\tilde{G}}} \right)
\end{equation}
where $m_{\tilde{g}}$ the gluino mass and $g$ the strong coupling. The gravitino yield is then given by 
\begin{equation} \label{}
\frac{dY_{\tilde{G}}}{dT}= -\frac{\left\langle \Sigma_{\tilde{G}}(T) v\right\rangle n_{rad}}{HT}\,,
\end{equation}
see e.g. \cite{Moroi:1993mb, Bolz:2000fu}. The  $H$ is the Hubble parameter and $n_{rad}=\zeta(3)T^3/\pi^2$. For Lightest Supersymmetric Particle (LSP) gravitino we can neglect the yield of the helicity $\pm 3/2$ component for temperatures less than 
\begin{equation}\label{T3/2}
T_{rh}< 10^{12} m^{-1}_{\tilde{G}}\, \text{GeV}^2\, .
\end{equation}
Whence, focusing on the interactions of the helicity $\pm 1/2$ modes the yield $Y_{\tilde{G}}$ is given by
\begin{equation} \label{int-9}
Y_{\tilde{G}}(T) = -\text{D} \left\{ \frac{n_{rad}\left\langle \Sigma^{(1/2)}_{\tilde{G}} v\right\rangle}{HT}\right\} \int^T_{T_{rh}}dT\, b^2_R(T)
\end{equation}
where $\Sigma^{(1/2)}_{\tilde{G}}\propto g^2m^2_{\tilde{g}}/(3m^2_{\tilde{G}}M^2_P)$ and D the dilution factor. We considered $Y_{\tilde{G}}(T_{rh}) \ll Y_{\tilde{G}}(T)$ i.e. the dominant source of gravitino production are the scatterings in the thermal plasma and any pre-inflationary abundance was diluted by inflation and 
\begin{equation}
\left\langle \Sigma^{(1/2)}_{\tilde{G}}(T) v\right\rangle =\left\langle \Sigma^{(1/2)}_{\tilde{G}}v \right\rangle b^2_R(T)
\end{equation}
as we can see from (\ref{crossT-9}). We note that the quantity in the brackets of (\ref{int-9}) has no explicit temperature dependence. 

For models that exhibit exact thermal restoration of the $U(1)_R$ symmetry the $b_R(T)$ can be approximated  by a step function: $b_R(T)=0$ for $T>T_0$ and $b_R(T)=1$ for $T<T_0$. Hence, the gravitino yield (\ref{int-9}) for $T_{rh}>T_0$ reads in this case:
\begin{equation} \label{intR-9}
\left. Y_{\tilde{G}}(T)
 \simeq \frac{g_{*s}(T)}{g_{*s}(T_{0})} \left\{ \frac{n_{rad}\left\langle \Sigma^{(1/2)}_{\tilde{G}} v\right\rangle}{HT} \right\}\right|_{T_{0}} T_0
\end{equation}
where we took into account that $T\ll T_0$. The ratio $g_{*s}(T)/g_{*s}(T_0)$ corresponds to the dilution factor where $g_{*s}$ the number of effectively massless degress of freedom \cite{Kolb:1990vq}. Therefore, for $T<1$ MeV, i.e. after nucleosynthesis, for a decoupled gravitino, $\mu\simeq 100$ GeV and for $T_{rh}>T_0$ the gravitino abundance is 
\begin{equation} \label{OR-9}
\Omega_{\tilde{G}}h^2 \simeq 0.2 \left(\frac{T_0}{10^{8}\,\text{GeV}} \right) \left(\frac{\text{GeV}}{m_{\tilde{G}}} \right) \left(\frac{m_{\tilde{g}}(\mu)}{1\, \text{TeV}} \right)^2\,
\end{equation}
i.e. it is controled by the $T_0$: the $R$-critical temperature given by the eq. (\ref{TR}).

For models that break explicitly the $R$-symmetry like those previously discussed  the $b_R(T)$ is given by (\ref{bRa})
and the relevant part of the integral (\ref{int-9}) is
\begin{equation} \label{apI}
\left. \int^T_{T_{rh}} dT\, b^2_R(T)= \frac{T_0}{2} \left\{\frac{T_0 T}{T^2_0+T^2} +\text{Arctan}\left(\frac{T}{T_0}\right) \right\} \right|^{T}_{T_{rh}}\,.
\end{equation}
For reheating temperatures larger than the $T_0$, which are the cases that we are interested in and especially for $T_{rh}\gg T_0$, the integral approximates to $-T_0\theta_{rh}/2$;
where, $\theta_{rh} \equiv \text{Arctan}(T_{rh}/T_0)$ a coefficient that takes values: $\pi/4<\theta_{rh} <\pi/2$.
For reheating temperatures lower than $T_0$  the integral (\ref{apI}) converges to $(-)T_{rh}$ value as expected. 

Therefore for $T_{rh}> T_0 $ the (\ref{int-9}) reads
\begin{equation} \label{Y}
\left. Y_{\tilde{G}}(T) \simeq \frac{g_{*s}(T)}{g_{*s}(T_{0})} \left\{ \frac{n_{rad}\left\langle \Sigma^{(1/2)}_{\tilde{G}} v\right\rangle}{HT}\right\}\right|_{T_{0}} \frac{T_0}{2}\theta_{rh}\
\end{equation}
where we considered the coefficients (i.e. the $g_{*s}$, $g_*$ factors \cite{Kolb:1990vq}) in front of the integral to be dominated by the value given at the temperature $T_0$. The resulting gravitino abundance is $\theta_{rh}/2$ times the (\ref{OR-9})
\begin{equation} \label{OgA}
\Omega_{\tilde{G}}h^2 \simeq 0.1 \left(\frac{\theta_{rh} T_0}{10^8\,\text{GeV}} \right) \left(\frac{\text{GeV}}{m_{\tilde{G}}} \right) \left(\frac{m_{\tilde{g}}(\mu)}{1\, \text{TeV}} \right)^2\, .
\end{equation}
The expression (\ref{OgA}) has a very mild dependence on the reheating temperature. We can say that the gravitino abundance is reheating-temperature-independent as long as the reheating temperature is larger than $T_0$ and not so large that the $3/2$ component to be overproduced. Indeed, for temperatures $T_{rh}>10\, T_0$ the $\theta_{rh}$ takes values only between 1.50 and $1.57$.  Substituting $\theta_{rh}\sim 1.5$ and  $T_0= 4F/(\sqrt{N}\lambda \Lambda)$ where $4F\simeq16.6\times10^{18} m_{\tilde{G}}\text{GeV}$ the expression (\ref{OgA}) attains the form
\begin{equation} \label{OgA2}
\Omega_{\tilde{G}}h^2 \sim 0.15 \times \frac{16.6}{\sqrt{N}} \left(\frac{10^{10} \, \text{GeV}}{\lambda\Lambda} \right) \left(\frac{m_{\tilde{g}}(\mu)}{1\, \text{TeV}} \right)^2\, .
\end{equation}
 
We mention, that in the case of approximate $R$-symmetric models the gravitino abundance is slightly smaller by a factor of $2/\theta_{rh}$ compared with the case of exact restoration of the $R$-symmetry. The expectation is that the gravitino production is stronger suppressed when an exact restoration of the $R$-symmetry takes place than when the restoration is approximate. This small discrepancy originates from the fact that we simplified the thermal evolution of the $X(T)$ by assuming a step-like behaviour while a second order phase transition is not a discontinuous process; instead at the critical temperature $T_0$ the barrier vanishes and the transition occurs smoothly. The $T_0$ value of the integral (\ref{intR-9})
and thus the $Y_{\tilde{G}}$ bound in that case is a conservative one.

\paragraph*{Discussion and Conclusions.}

The expressions (\ref{OR-9}) and (\ref{OgA}),(\ref{OgA2}) are a remarkable result. Firstly, the gravitino abundance exhibits an insensitivity to the gravitino mass and the reheating temperature. Secondly, the quantities which control the yield are the supersymmetry breaking fundamental parameters $\lambda$ and $\Lambda$. For $\sqrt{N}={\cal O}(1-7)$ \cite{Giudice:1998bp} and $m_{\tilde{g}}\sim 1$ TeV, the gravitino does not overclose the universe if
\begin{equation}
\lambda \Lambda \gtrsim 10^{10} \, \text{GeV}.
\end{equation}
When the lower bound is saturated it can account for the dominant dark matter component; 
 for instance, when $\lambda \sim 10^{-5}$ and $\Lambda= {\cal O}$(GUT) scale, the gravitino is the dark matter of the universe. It is interesting to note 
that a GUT scale cut-off  is expected in many theories like those discussed here. This fact also implies the smallness of the coupling since in the IR theory the coupling $\lambda$ is expected to be suppressed, see e.g.  \cite{Murayama:2007fe}.

In addition, considering that the supersymmetry breaking local minimum must be thermally preferred we conclude that small values for the coupling $\lambda\lesssim 10^{-4}$ are cosmologically favourable \cite{Dalianis:2010pq}. On the contrary, the increase of the coupling $\lambda$ value renders the supersymmetry preserving 
vacua more attractive.
We thus find an interesting window of values where the supersymmetry breaking vacuum is thermally  selected and the gravitino (nearly automatically) does not overclose the universe, or even accounts for the dominant dark matter component. This parameter space also specifies the gravitino mass range. 

The interactions of the helicity $\pm 3/2$ that are $T^2/M^2_P$ suppressed impose an upper bound on the $T_{rh}$.  According to the relation (\ref{T3/2}) and for $\lambda \Lambda \gtrsim 10^{10}$ GeV, a gravitino, for instance,  with mass $1$ GeV (100 MeV) constrains the reheating temperature to be $T_{rh} \lesssim 10^{12}$ GeV ($10^{13}$ GeV) which is about $10^4$ ($10^6$)  times relaxed relatively to the conventional bound on the reheating temperature for the stable gravitino. For $m_{\tilde{G}}\lesssim1 $ MeV the yield of the $3/2$ gravitino component can be safely neglected \cite{Enqvist:1993fm} and thus the total yield is given solely by the (\ref{intR-9}) or (\ref{Y}). An interesting remark is that when the conventional (i.e. ignoring the $b_R(T)$ factor) freeze-out gravitino temperature 
is larger that the $T_0$ the light gravitino never actually equilibrates. In other words, in this case there is \itshape no \normalfont freeze-out temperaure for the gravitinos.

The $R$-symmetry is generally expected to be restored if large reheating temperatures were realized i.e. $T_{rh}>T_0$. In the minimum of the finite temperature effective potential the entropy maximizes. 
 Here, the thermalized degrees of freedom are expected to be the (M)SSM degrees of freedom and the messenger fields. The later have tree level mass $\sim \lambda X$ and become light at the origin.
Thermalized hidden sector fields that become light at the $X=X_0$ could render the $R$-breaking minimum thermally favourable rather than the origin $X=0$. However, this requires $R$-violating interactions. Furthermore, a thermalized hidden sectror that couples directly to the Goldstino would lead to a large $dn_{\tilde{G}}/dt$ with a Goldstino production rate scaling like $T^8/F^2$ \cite{Leigh:1995jw}. If such a hidden sector exists it must be nearly unpopulated.  On the other hand, the messengers although thermalized give a $\lambda^4$ suppressed Goldstino production rate that can be neglected since $\lambda \ll 1$. These arguments indicate that  
an $R$-symmetry restoration probably takes place in a high temperature environment of the early universe.

The $m_X\sim F/\Lambda$ is of the order of the electroweak scale thereby, it is the smallness of the Yukawa coupling at the messenger superpotential that makes the $T_0$ large suggesting a ${\cal O}$(MeV)-${\cal O}$(GeV) mass range gravitino for dark matter (\ref{OgA}).

The conclusion is that the gravitino cosmological problem can be actually absent in the most general class of gauge mediation models without including additional ingredients or assumptions. This is deduced from the expected gaugino mass $R$-suppression described by the eq. (\ref{ratio-9}). Moreover, much like the neutralino LSP which has a fixed thermal abundance as long as the $T_{rh}$ is larger than the freeze-out temperature, so the gravitino LSP has a relic abundance fairly insensitive to high reheating temperatures as long as $T_{rh}>T_0$. Barring that this fact is good news for thermal leptogenesis we emphasize that the gravitino can be indeed considered a compelling candidate for the bulk dark matter of the universe.

The author would like to thank W. Buchm\"uller, Z. Lalak and K. Turzynski for discussions. This work was partially supported by the 
EC 6th FrameworkProgramme MRTN-CT-2006-035863 and by the Polish
Ministry for Science and Education under grant N N202 091839.

\end{document}